\documentclass[twocolumn,showkeys,preprintnumbers,amsmath,amssymb]{revtex4}

\usepackage{graphicx}
\usepackage{dcolumn}
\usepackage{bm}

\begin{document}

\title{Carmeli's accelerating universe is spatially flat without dark matter}

\author{John G. Hartnett}
 \email{john@physics.uwa.edu.au}
\affiliation{School of Physics, the University of Western Australia\\35 Stirling Hwy, Crawley 6009 WA Australia}%


\date{\today}

\begin{abstract}
Carmeli's 5D brane cosmology has been applied to the expanding accelerating universe and it has been found that the distance redshift relation will fit the data of the high-z supernova teams without the need for dark matter. Also the vacuum energy contribution to gravity,  $\Omega_{ \Lambda}$ indicates that the universe is asymptotically expanding towards a spatially flat state, where the total mass energy density $\Omega + \Omega_{ \Lambda} \rightarrow 1$. 
\end{abstract}


\maketitle

\section{Introduction}
The Carmeli cosmology \cite{Behar2000, Carmeli2002} is as revolutionary in its implementation as it is in its interpretation. The metric used by Carmeli is unique in that it extends the number of dimensions of the universe by, either one dimension if we consider only the radial velocity of the galaxies in the Hubble flow or by three if we consider all three velocity components. We will confine the discussion in this paper to only one extra dimension as does Carmeli. In that case the line element in five dimensions becomes
\begin{equation} \label{eqn:metric}
ds^{2}=(1+\frac{\Phi}{c^{2}})c^{2}dt^{2}-dr^{2}+(1+\frac{\Psi}{\tau^{2}})\tau^{2}dv^{2},
\end{equation}
where $ dr^{2}=(dx^1)^{2}+(dx^2)^{2}+(dx^3)^{2}$ and $\Phi$ are $\Psi$ potential functions to be determined. The time ($t$) is measured in the observer's frame. The new dimension ($v$) is the radial velocity of the galaxies in the expanding universe, in accordance with Hubble flow. The parameter $\tau$, the Hubble-Carmeli constant, is a constant at any epoch and its reciprocal (designated $h$) is approximately the Hubble constant $H_{0}$. 

The line element represents a spherically symmetric isotropic universe, and the expansion is the result of \textit{spacevelocity} expansion. The expansion is observed at a definite time and thus $dt = 0$. Taking into account $ d\theta = d\phi = 0 $ (isotropy condition) and  equation  (\ref {eqn:metric})  becomes
\begin{equation} \label{eqn:phasespacemetric}
-dr^{2}+(1+\frac{\Psi}{\tau^{2}})\tau^{2}dv^{2}=0.
\end{equation}

\section{Phase space equation}
The solution of (\ref {eqn:phasespacemetric}) (given by equation B.38 and solved in section B.10 in \cite{Carmeli2002}) is reproduced here.

\begin{equation} \label{eqn:phasespacederiv}
\frac{dr}{dv}= \tau \sqrt{ 1+(1-\Omega) \frac{r^2}{c^2 \tau^2}}
\end{equation}
\\
The parameter $\Omega$ is the mass/energy density of the universe expressed as a fraction of the critical or  ``closure'' density, i.e. $\Omega = \rho_m/\rho_c$, where $\rho_m$ is the averaged matter/energy density of the universe. In this model, \begin{displaymath}
\rho_c= \frac{3}{8 \pi G \tau^{2}}= 10^{-29} \; g \; cm^{-3}.
\end{displaymath}
\\
Then (\ref {eqn:phasespacederiv}) may be integrated exactly to get 

\begin{equation} \label{eqn:phasespacesoln}
r(v)= \frac{c \tau}{\sqrt{1-\Omega}}\sinh \left( \frac{v}{c} \sqrt{1-\Omega } \right)
\qquad \forall \Omega.
\end{equation}
\\
Carmeli expanded (\ref {eqn:phasespacesoln}) in the limit of small $z = v/c$ and small $\Omega$ to get
\begin{equation} \label{eqn:phasespacesolnC}
r =  \tau v \left(1+ (1- \Omega)\frac{v^2}{6 c^2}  \right)
\end{equation}
                  
\begin{equation} \label{eqn:phasespacesolnCz}
\Rightarrow \frac {r} {c \tau} = z  \left(1+ (1- \Omega)\frac{z^2}{6}  \right)
\qquad  \forall \Omega <1, z < 1.
\end{equation}                  

Thus we can write the expansion in terms of normalized or natural units $ r/c \tau $. Equation (\ref {eqn:phasespacesolnCz}) is plotted in fig. \ref{fig:fig1} for various values of $\Omega = 1, 0.24 $ and $ 0.03$. Let us now re-write (\ref {eqn:phasespacesoln}) in terms of natural units and for small $z$ but arbitrary $\Omega$, and we get
\begin{equation} \label{eqn:phasespacesolnnatural}
\frac {r} {c \tau}= \frac {\sinh (z \sqrt{1-\Omega})} {\sqrt{1-\Omega}}.
\end{equation}
\\
Equation (\ref {eqn:phasespacesolnnatural}) produces curves almost indistinguishable from (\ref {eqn:phasespacesolnCz}) so this verifies that the approximations work well for $z < 1$. 

\section{Density verses redshift}
\begin{figure}
\includegraphics[width = 3.5 in]{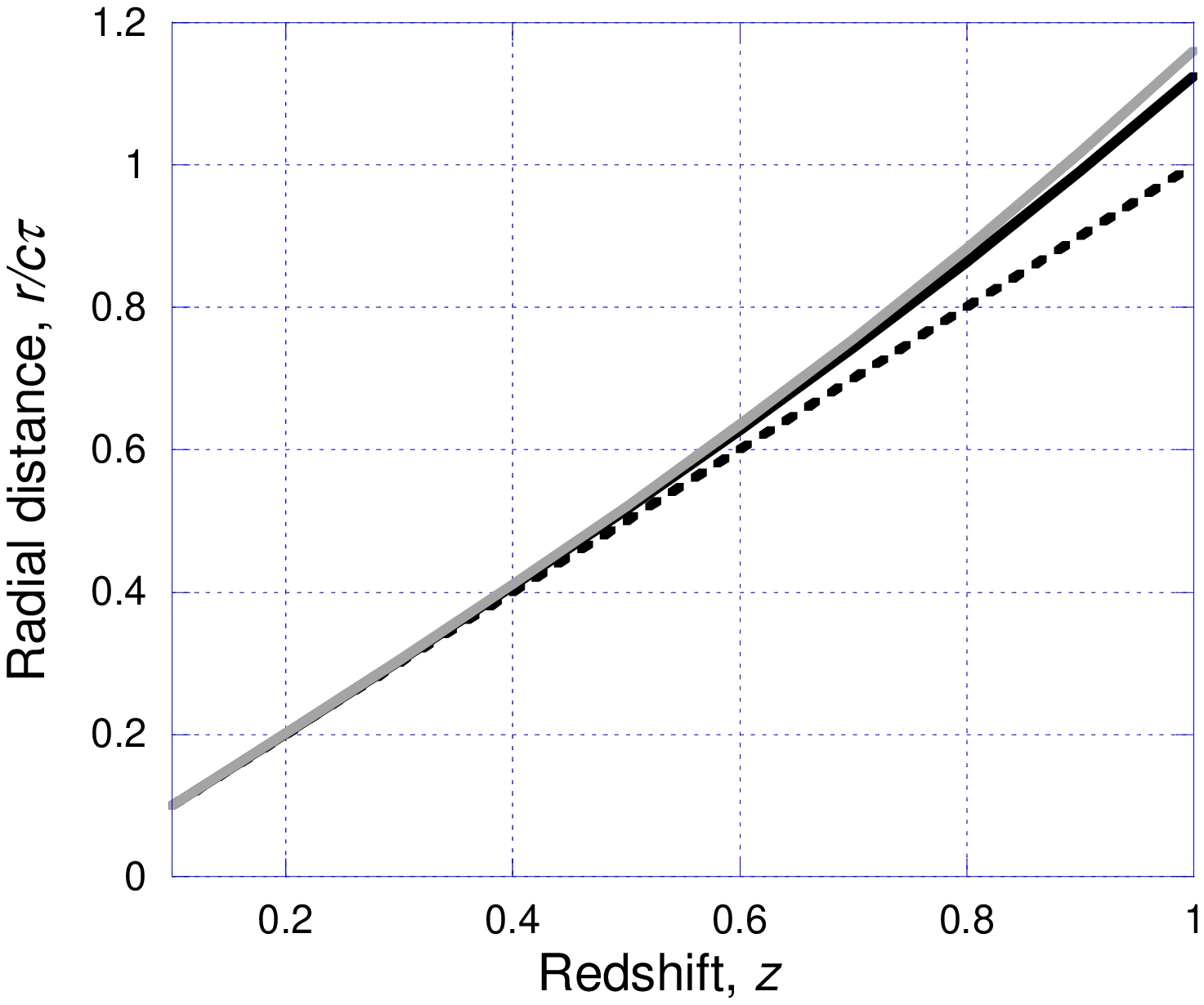}
\caption{\label{fig:fig1} Plot of (\ref {eqn:phasespacesolnCz}), $r/c\tau$   vs redshift ($z$) for $\Omega = 1$ (broken line), $\Omega = 0.245$ (solid black line) and $\Omega = 0.03$ (solid grey line)}
\end{figure}

Now let us consider what happens to the density of matter as we look back in the cosmos with redshift, $z$. It was assumed in fig. \ref{fig:fig1} that the value of $\Omega$ is fixed for each curve. Carmeli does this also in figure A4, page 134 in \cite{Carmeli2002}.  However, more correctly $\Omega$ varies as a function of $z$. For flat space we assume the following relation to hold, 
\begin{equation} \label{eqn:densityeqn}
\frac {\rho_{m}} {\rho_{0}} = (1+z)^3 = \frac {\Omega}{\Omega_{0}},
\end{equation}                           
where $\rho_{m}$ is a function of the redshift $z$, and $\rho_{0}$ is the averaged matter density of the universe locally ($z \approx 0$). The parameter $\Omega_{0}$ is then the local averaged matter density expressed as a fraction of ``closure'' density. Equation (\ref {eqn:densityeqn}) results from the fact that as the redshift increases the volume decreases as $(1 + z)^{3}$. Notice at $z = 1$ that the universe is 8 times smaller in volume and therefore 8 times more dense, that is, at $z =1$,  $\Omega = 8 \Omega_{0}$. 

\begin{figure}
\includegraphics[width = 3.5 in]{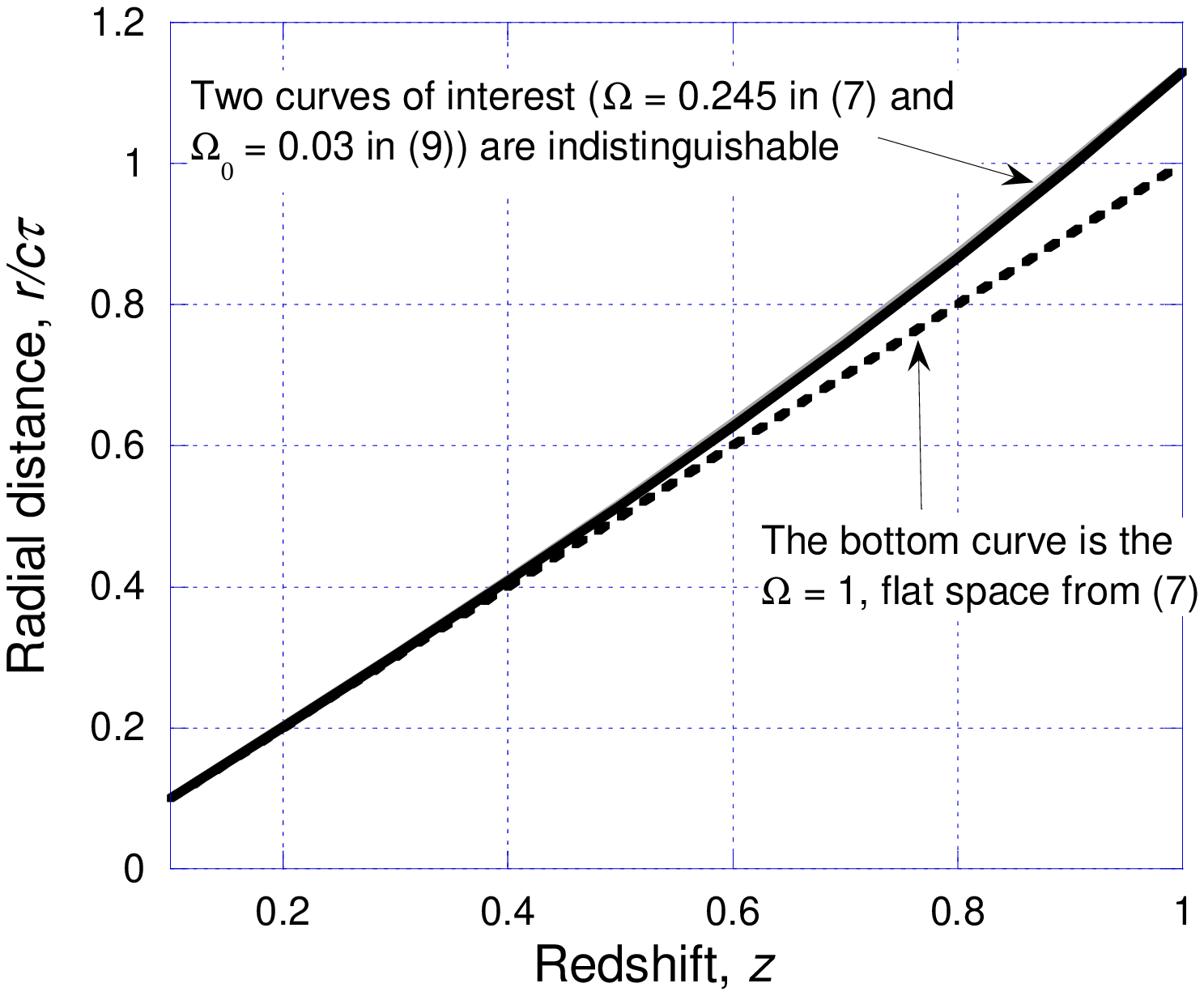}
\caption{\label{fig:fig2} Plot of (\ref {eqn:phasespacesolnnatural}) with $\Omega = 1$ (broken line) and $\Omega = 0.245$ (solid black line) and (\ref {eqn:phasespacesolnnaturald}) with $\Omega_{0} = 0.03$ (solid grey line). Note: the top two curves lay on top of each other}
\end{figure}

Substituting (\ref {eqn:densityeqn}) into (\ref {eqn:phasespacesolnnatural}) we get 
\begin{equation} \label{eqn:phasespacesolnnaturald}
\frac {r} {c \tau}= \frac {\sinh \left(z \sqrt{1-\Omega_{0}(1+z)^3 }\right)} {\sqrt{1-\Omega_{0}(1+z)^3}}.
\end{equation}
\\
Carmeli was able to simulate the form of the $0.1 < z < 1$ redshift data of \cite{Riess1998} published in 1998 which announced an accelerating universe following the observations of \cite{Garnavich1997, Perlmutter1997}. See figure A4, page 134 in \cite{Carmeli2002}. But in fact he had predicted this in 1996 \cite{Carmeli1996}. So this means that Carmeli assumed a value of total matter (normal + dark matter) density $\Omega = 0.245$, which was the accepted value in 1998.

Now let's plot (\ref {eqn:phasespacesolnnatural}) with $\Omega = 0.245$ and (\ref {eqn:phasespacesolnnaturald}) with $\Omega_{0} = 0.03$. See fig. \ref{fig:fig2}. This means that my modified equation (\ref {eqn:phasespacesolnnaturald}) with $\Omega_{0} = 0.03$ gives the same result as Carmeli's unapproximated equation (\ref {eqn:phasespacesolnnatural}) with his assumed value of  $\Omega = 0.245$, but that included dark matter. In fact, comparing (\ref {eqn:phasespacesolnnatural}) and (\ref {eqn:phasespacesolnnaturald}), a local matter density of only $\Omega_{0} =$ 0.03--0.04 is necessary to have agreement. This effectively \textit{eliminates the need for the existence of dark matter} on the cosmic scale.

Table I shows the critical data from the comparison at redshifts between $z = 0.25$ and $z = 1$. It can be seen that the difference between the two equations over the domain of the measurements is much less significant than the fit to the data.  If we assume  $\Omega_{0} = 0.04$ instead of  $\Omega_{0}  = 0.03$, since both are within measured parameters, we get closer agreement at smaller redshifts but worse near $z = 1$. \\

\begin{figure}
\includegraphics[width = 3.5 in]{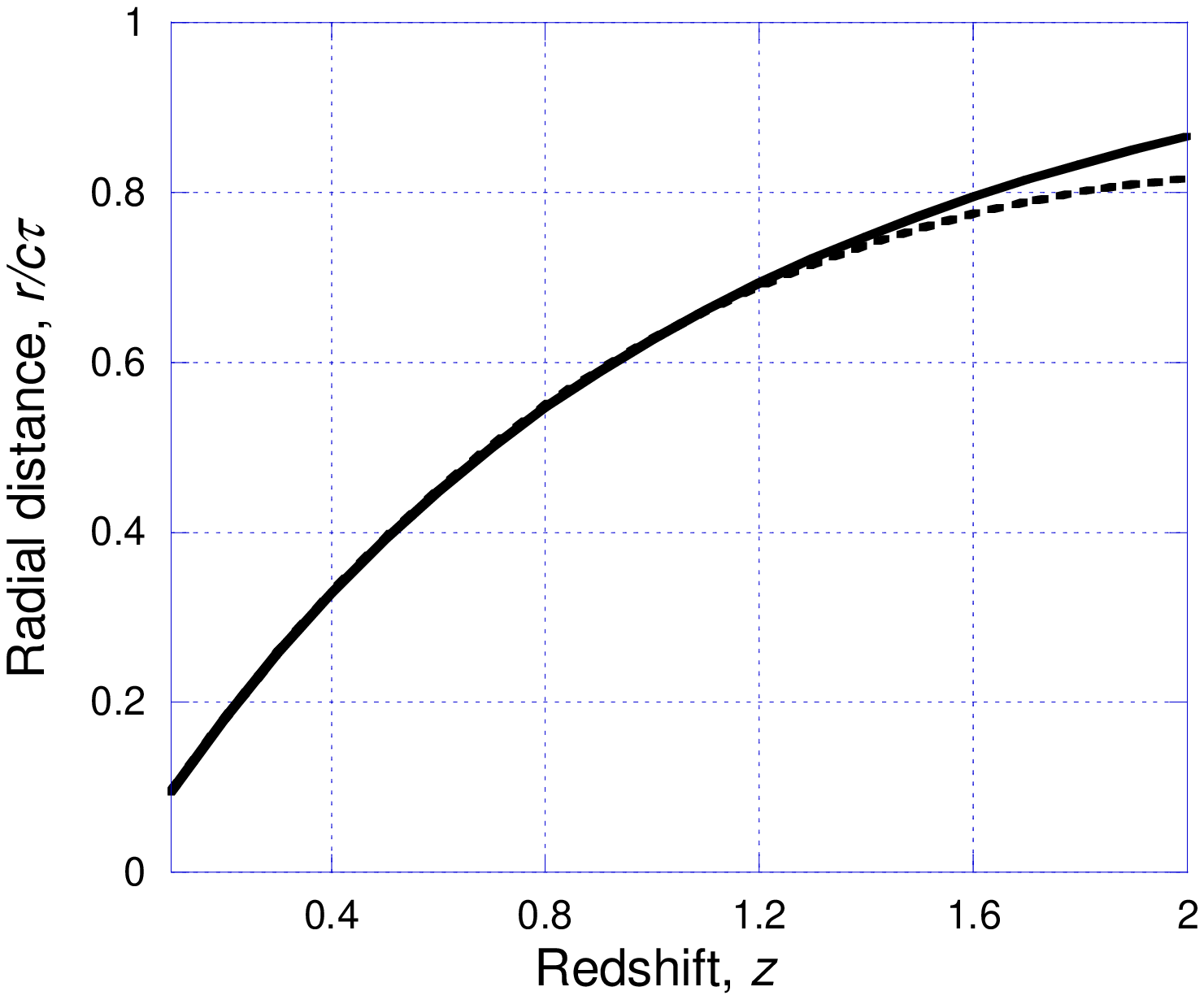}
\caption{\label{fig:fig3} Plot of (\ref {eqn:phasespacesolnnaturalr}) with $\Omega = 0.245$ (solid curve) and (\ref {eqn:phasespacesolnnaturalrd}) with $\Omega_{0} = 0.03$ (broken curve). Note: the two curves separate for $z > 1.2$}
\end{figure}

In any case (\ref {eqn:phasespacesolnnatural}) and (\ref {eqn:phasespacesolnnaturald}) must be modified as $z \rightarrow 1$ to allow for relativistic effects, by replacing $v/c$ with the relativistic form $\varsigma = v/c = ((1+z)^2-1)/((1+z)^2+1)$. Therefore we can re-write (\ref {eqn:phasespacesolnnatural}) and (\ref {eqn:phasespacesolnnaturald}) respectively as 
\begin{equation} \label{eqn:phasespacesolnnaturalr}
\frac {r} {c \tau}= \frac {\sinh \left(\varsigma \sqrt{1-\Omega}\right)}{\sqrt{1-\Omega}} 
\end{equation}
and
\begin{equation} \label{eqn:phasespacesolnnaturalrd}
\frac {r} {c \tau}=\frac {\sinh \left(\varsigma \sqrt{1-\Omega_{0}(1+z)^3 }\right)}{\sqrt{1-\Omega_{0}(1+z)^3}} ,
\end{equation}
where the varying matter density has been taken into account. In fig. \ref{fig:fig3}, (\ref {eqn:phasespacesolnnaturalr}) and (\ref {eqn:phasespacesolnnaturalrd}) are compared. The density approximation may be no longer valid past $z = 1$, because it is shown below that the vacuum energy term dominates and the universe is far from spatially flat.

\begin{table*}
\begin{center}
\caption{\label{tab:table1}Comparison of equations (\ref {eqn:phasespacesolnnatural}) and (\ref {eqn:phasespacesolnnaturald})}
\begin{tabular}[c]{|c|cccc|}
\hline
\textbf{Redshift z}  & 0.25 & 0.5 & 0.75 & 1.0\\
\hline
$\mathbf{r/c\tau}$ \textbf{from (\ref {eqn:phasespacesolnnatural}) with} $\mathbf{\Omega  = 0.245}$ & $0.251984$ & $0.515984$ & $0.804591$ & $1.13157$\\
$\mathbf{r/c\tau}$  \textbf{from (\ref {eqn:phasespacesolnnaturald}) with} $\mathbf{\Omega_{0} = 0.03}$  & $0.252459$ & $0.518935$ & $0.810416$ & $1.13157$\\
\textbf{\% difference with} $\mathbf{\Omega_{0} = 0.03}$  & $0.19$ &	$0.57$ &	$0.72$ &	$0.00$\\
\textbf{\% difference with} $\mathbf{\Omega_{0} = 0.04}$  & $0.17$ &	$0.43$ &	$0.23$ &	$1.28$\\
\hline
\end{tabular}
\end{center}
\end{table*}

\section{Hubble parameter}
Based on the above analysis we can rewrite equation A.54 from \cite{Carmeli2002} for $H_{0}$  as

\begin{equation} \label{eqn:Hubbleparam}
H_{0} =  h\left(1- (1- \Omega_{0}(1+z)^3)\frac{z^2}{6}  \right) \; \forall \Omega,z < 1, 
\end{equation}                  
which according to equation A.51 from \cite{Carmeli2002} may be further generalized without approximation, and using the relativistic form of the redshift. Still that equation may only be approximate for $z > 1$ because of the density assumptions. However it becomes 

\begin{figure}
\includegraphics[width = 3.5 in]{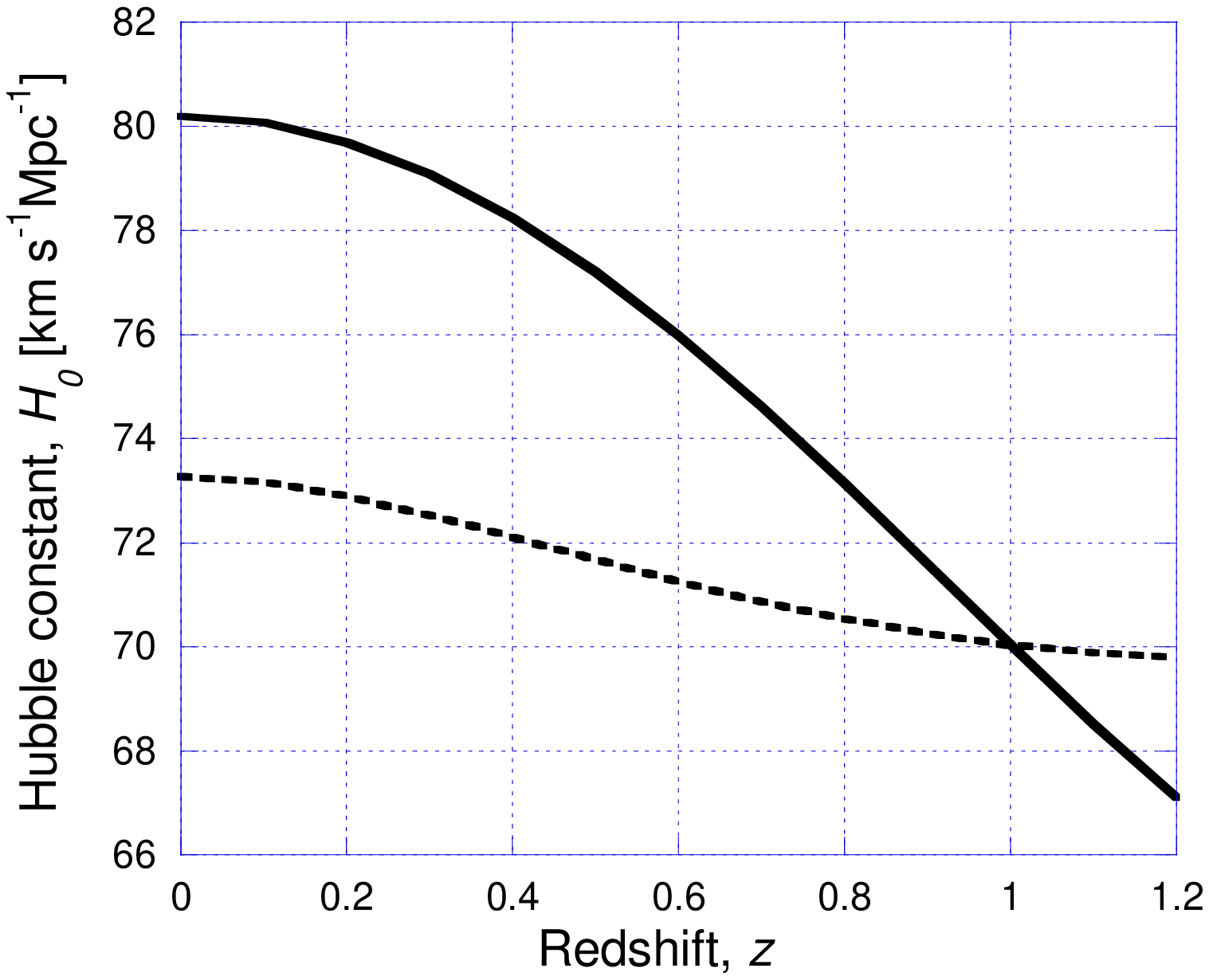}
\caption{\label{fig:fig4} Plot of (\ref {eqn:Hubbleparam}) (solid curve) and (\ref {eqn:Hubbleparamgen}) (broken curve). Note: the two curves intersect at $z = 1$ where $H_{0} = 70 \; km \; s^{-1}\; Mpc^{-1}$}
\end{figure}

\begin{equation} \label{eqn:Hubbleparamgen}
H_{0} =  h \frac{\xi} {\sinh \xi},
\end{equation}                 
where $\xi = \frac {(1+z)^2-1} {(1+z)^2+1} \sqrt{1-\Omega_{0}(1+z)^3} $.

Both (\ref {eqn:Hubbleparam}) and (\ref {eqn:Hubbleparamgen}) have been plotted in fig. \ref{fig:fig4}, and for Carmeli's chosen value of $H_{0} \approx 70 \; km \; s^{-1} \; Mpc^{-1}$ at $z = 1$ in (\ref {eqn:Hubbleparam}) yields $h \approx 80.2 \; km \; s^{-1} \; Mpc^{-1}$ (very close to Carmeli's value) but (\ref {eqn:Hubbleparamgen}) yields $h \approx 73.27 \; km \; s^{-1} \; Mpc^{-1}$. This means without the small z approximation the value of $h$ is reduced when compared to that in \cite{Carmeli2002}.

\section{Dark energy}
The vacuum or dark energy parameter $\Omega_{\Lambda}$ does not appear explicitly in Carmeli's model. It is only by a comparison with F-L models can an assignment be made. On page 138 of  \cite{Carmeli2002} by comparing with the standard model it is shown that $\Omega_{\Lambda} = (H_{0}/h)^2$, therefore we can write

\begin{figure}
\includegraphics[width = 3.5 in]{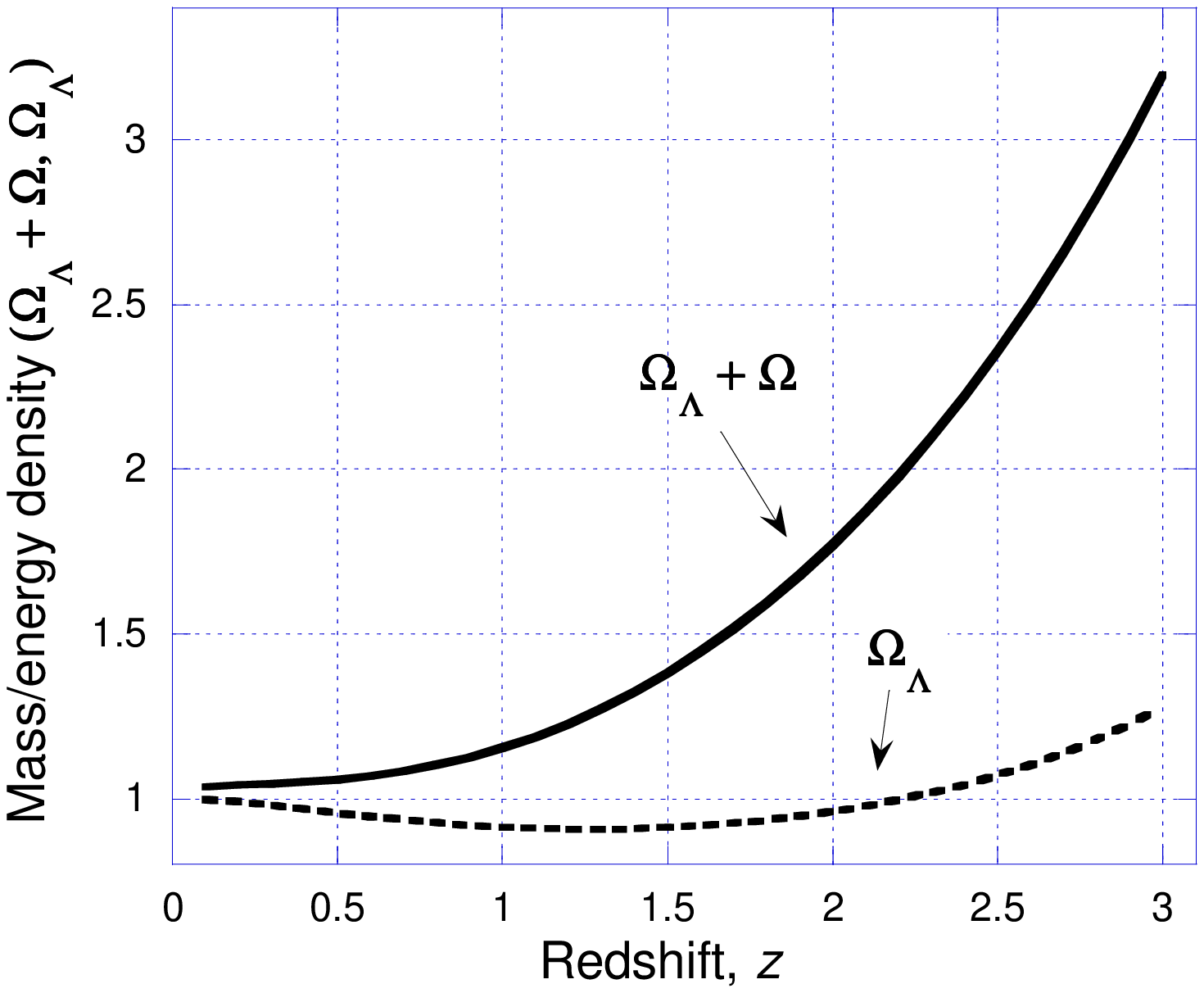}
\caption{\label{fig:fig5} Plot of  $\Omega_{\Lambda}$ (broken curve) and total density  $\Omega + \Omega_{\Lambda}$  (solid curve) as a function of redshift $z$. Notice that  $\Omega_{\Lambda}$ tends to unity as $z$ tends to zero and the total density tends to the local matter density $\Omega_{0}$ plus the vacuum energy density $\Omega_{\Lambda}$}
\end{figure}

\begin{equation} \label{eqn:Darkenergy}
\Omega_{\Lambda} =   \left(\frac{\xi} {\sinh \xi}\right)^2.
\end{equation} 
From (\ref{eqn:Darkenergy}) it is expected that using the unapproximated equation (\ref{eqn:Hubbleparamgen}) for $H_{0}$ the value of $\Omega_{\Lambda}$ will be larger than Carmeli's value using the form of (\ref{eqn:phasespacesolnCz}).  Fig. \ref{fig:fig5} shows the values for the vacuum energy density  $\Omega_{\Lambda}$ (broken curve) and for the total energy density  $\Omega + \Omega_{\Lambda}$   (solid curve) as a function of redshift, $z$. From (\ref{eqn:Darkenergy}) it follows that as the universe expands the total density tends to the vacuum energy density  $\Omega_{\Lambda}\rightarrow 1$ (since $\Omega_{0}\rightarrow 0$). This means a totally 3D spatially flat universe in a totally relaxed state. 

For small $z$ the total density becomes

\begin{equation} \label{eqn:Darkenergysmz}
\Omega + \Omega_{\Lambda} \approx   (1+\Omega_{0})+3z\Omega_{0}.
\end{equation} 
It follows from (\ref {eqn:Darkenergysmz}) that for  $\Omega_{0} = 0.03$ at z = 0 the total density  $\Omega + \Omega_{\Lambda}\approx 1.03$. This value is consistent with Carmeli's result of $1.009$.  However, it follows from (\ref {eqn:densityeqn}) and (\ref {eqn:Darkenergysmz}) that the universe will always be open,  $\Omega < 1$ as it expands. From fig. \ref{fig:fig5} the total density  $\Omega + \Omega_{\Lambda}$  is always greater than unity and as the universe expands, it asymptotically approaches unity---therefore a spatially flat universe devoid of dark matter.

\section{Conclusion}
The 5D brane world of Moshe Carmeli has been has been applied to the expanding accelerating universe and the redshift distance relation has been generalised for redshifts up to at least $z = 1.2$.  It has been found that if a certain form of the dependence of baryonic matter density on redshift is assumed then the resulting distance-redshift relation will approximate the form of the data of the high-$z$ supernova teams without the need for dark matter.  

Even though it does not explicitly appear in the Carmeli \emph{spacevelocity} metric, the vacuum energy contribution to gravity, $\Omega_{\Lambda}$ tends to unity as a function of decreasing redshift.  Also since the baryonic matter density  $\Omega_{0}\rightarrow 0$ as the universe expands, the total mass/energy density $\Omega + \Omega_{\Lambda}\rightarrow 1$. This indicates that the universe, though always open because $\Omega < 1$, is asymptotically expanding towards a spatially flat state.
\\
\section{Acknowledgment}
I would like to thank Prof. Moshe Carmeli for many valuable discussions.\\


\begin{thebibliography}{99}
\bibitem{Carmeli1996} Carmeli, M. (1996). \emph{Cosmological general relativity}, Commun. Theor. Phys. \textbf{5}:159
\bibitem{Behar2000} Behar, S. Carmeli, M. (2000) \emph{Cosmological Relativity: A New Theory of Cosmology} Int. J. Theor. Phys.  \textbf{39}(5)
\bibitem{Carmeli2002} Carmeli, M. (2002). \emph{Cosmological Special Relativity}. Singapore, World Scientific. 
\bibitem{Garnavich1997} Garnavich, P. M., \textit{et al}. (1997). \emph{Constraints on Cosmological Models from Hubble Space Telescope Observations of High-$z$ Supernovae} Bulletin of the American Astronomical Society \textbf{29}(7): 1350
\bibitem{Perlmutter1997} Perlmutter, S., \textit{et al}. (1997). \emph{Cosmology From Type Ia Supernovae: Measurements, Calibration Techniques, and Implications.} Bulletin of the American Astronomical Society \textbf{29}(5): 1351
\bibitem{Riess1998} Riess, A. G., \textit{et al}. (1998). \emph{Observational evidence from supernovae for an accelerating universe and a cosmological constant} Astron. J. \textbf{116}(Sept): 1009-1038

\end{thebibliography}
\end{document}